# Evidence for Primordial Black Hole Final Evaporation: Swift, BATSE and KONUS and Comparisons of VSGRBs and Observations of VSB That Have PBH Time Signatures

David B. Cline and Stan Otwinowski
UCLA Department of Physics & Astronomy
405 Hilgard Avenue, Los Angeles CA 90095

**Abstract**

More than a decade ago we identified a class of VSGRB ($T_{90}$ < 100 ms) as having unusual properties: (1) galactic position asymmetry, (2) very hard $\gamma$ spectrum, (3) possible evidence for galactic origin of these events. We now study the recent Swift data and show that a VSGRB enhancement consistent BATSE and KONUS exists. We estimate that this is now a total 4.5$\sigma$ observation. We then study the VSB for evidence of the time structure expected for PBH evaporation. Several of the events show the general time structure expected for PBH evaporation. If correct, then PBH must exist in this galaxy. Since even large detectors like BATSE record only a few VSB per year the density of PBH can still be very small and it is hard to predict a rate for the Fermi spacecraft LAT.

## 1. Introduction

For many years our group has studied the short time duration gamma ray bursts, looking for evidence of primordial black hole evaporation. We have identified an enhancement for $T_{90}$ time of less than about 100 ms time duration two long duration space experiment BATSE and KONUS and published the results [1][2][3][4][5][6]. Others have also noticed some of these effects [7]. One of the earliest predictions for PBH is in Ref. 8.

We now explore the hypothesis that a similar enhancement exists in the Swift spacecraft data in this paper. The net effect in all three data sets is a 4.5$\sigma$ effect now.

We also show that the VSB have a time structure distribution nearly exactly the shape expected for the final evaporation of a PBH (see Table 1). This follows the work of Hawking and others [9][10][11][12][13].

We now explore the possibility that this effect could be due to PBH evaporation at the end of the life of an evaporating PBH. We show that the size of the enhancement would be due to PBH evaporation in the final 100 ms of the lifetime and we now predict such an effect will be observed by the Fermi spacecraft LAT detector! Such evaporating PBH give a natural explanation of the T < 100 ms effects we have observed and published. The Swift data are key additions to this data set!

## 2. Explanation of the Enhancement of VSGRB for $T_{90}$ < 100 ms from the Final Evaporation of PBH

We display the full life history and parameters of a PBH evaporation in Table 1 using the formula from Hawking [9][10][11][12][13].

Table 1. Table 1. History of PBH evaporation. $M_{BH}$[g] – PBH mass, $t_{BH}$ – time PBH before final evaporation in seconds and years, TBH – PBH temperature in Kelvin, eV and GeV, RS[m] – PBH Schwarzschild radius in m, $L_{BH}$ – PBH luminosity ($L_{BH}$) in g/s and erg/s.

| $M_{BH}$[g] | $t_{BH}$[s] | $t_{BH}$[y] | $T_{BH}$[K] | $T_{BH}$[eV] | $T_{BH}$[GeV] | $R_s$[m] | $L_{BH}$[g/s] | $L_{BH}$[erg/s] |
|---|---|---|---|---|---|---|---|---|
| 1,00E+15 | 8,38E+19 | 2,66E+12 | 1,23E+11 | 1,06E+07 | 1,06E-02 | 1,48E-09 | 6,82E-06 | 6,13E+18 |
| 5,00E+14 | 1,05E+19 | 3,32E+11 | 2,46E+11 | 2,12E+07 | 2,12E-02 | 7,40E-10 | 3,85E-05 | 3,46E+19 |
| 1,00E+14 | 8,38E+16 | 2,66E+09 | 1,23E+12 | 1,06E+08 | 1,06E-01 | 1,48E-10 | 1,08E-03 | 9,67E+20 |
| 1,00E+13 | 8,38E+13 | 2,66E+06 | 1,23E+13 | 1,06E+09 | 1,06E+00 | 1,48E-11 | 1,08E-01 | 9,67E+22 |
| 1,00E+12 | 8,38E+10 | 2,66E+03 | 1,23E+14 | 1,06E+10 | 1,06E+01 | 1,48E-12 | 1,08E+01 | 9,67E+24 |
| 1,00E+11 | 8,38E+07 | 2,66E+00 | 1,23E+15 | 1,06E+11 | 1,06E+02 | 1,48E-13 | 1,08E+03 | 9,67E+26 |
| 1,00E+10 | 8,38E+04 | | 1,23E+16 | 1,06E+12 | 1,06E+03 | 1,48E-14 | 1,08E+05 | 9,67E+28 |
| 1,00E+09 | 8,38E+01 | | 1,23E+17 | 1,06E+13 | 1,06E+04 | 1,48E-15 | 1,08E+07 | 9,67E+30 |
| **1,00E+08** | **8,38E-02** | | **1,23E+18** | **1,06E+14** | **1,06E+05** | **1,48E-16** | **1,08E+09** | **9,67E+32** |
| 1,00E+07 | 8,38E-05 | | 1,23E+19 | 1,06E+15 | 1,06E+06 | 1,48E-17 | 1,08E+11 | 9,67E+34 |
| 1,00E+06 | 8,38E-08 | | 1,23E+20 | 1,06E+16 | 1,06E+07 | 1,48E-18 | 1,08E+13 | 9,67E+36 |
| 1,00E+05 | 8,38E-11 | | 1,23E+21 | 1,06E+17 | 1,06E+08 | 1,48E-19 | 1,08E+15 | 9,67E+38 |
| 1,00E+04 | 8,38E-14 | | 1,23E+22 | 1,06E+18 | 1,06E+09 | 1,48E-20 | 1,08E+17 | 9,67E+40 |
| 1,00E+03 | 8,38E-17 | | 1,23E+23 | 1,06E+19 | 1,06E+10 | 1,48E-21 | 1,08E+19 | 9,67E+42 |
| 1,00E+02 | 8,38E-20 | | 1,23E+24 | 1,06E+20 | 1,06E+11 | 1,48E-22 | 1,08E+21 | 9,67E+44 |
| 1,00E+01 | 8,38E-23 | | 1,23E+25 | 1,06E+21 | 1,06E+12 | 1,48E-23 | 1,08E+23 | 9,67E+46 |
| 1,00E+00 | 8,38E-26 | | 1,23E+26 | 1,06E+22 | 1,06E+13 | 1,48E-24 | 1,08E+25 | 9,67E+48 |
| 1,00E-01 | 8,38E-29 | | 1,23E+27 | 1,06E+23 | 1,06E+14 | 1,48E-25 | 1,08E+27 | 9,67E+50 |
| 1,00E-02 | 8,38E-32 | | 1,23E+28 | 1,06E+24 | 1,06E+15 | 1,48E-26 | 1,08E+29 | 9,67E+52 |
| 1,00E-03 | 8,38E-35 | | 1,23E+29 | 1,06E+25 | 1,06E+16 | 1,48E-27 | | |

Only the survival time of 1 sec to 1 ms would likely allow for a detectable GRB with existing space detectors. The energy of observation of a GRB goes (approximately) like the $T^2_{PBH}$ and increases rapidly. From Table 1 we find the approximate distance that BATSE could have detected the photons from the PBH and report these very approximate values in Table 2.

Table 2. Very Approximate Distance for PBH Detection by BATSE

| $T_{PBH}$ | $T_{90}$ | $V_{detector\ range}$ |
|---|---|---|
| ~4TeV | 1 sec | $[1]^3$ pc$^3$ |
| ~10TeV | 100ms | $[3]^3$ pc$^3$ |
| ~15TeV | 30ms | $[5]^3$ pc$^3$ |
| ~20TeV | 10ms | $[7]^3$ pc$^3$ |

Since the distance scales like d the detection rate scales like $[d]^3$. For example, for a $T_{90} = 1$ sec to $T_{90} = 10$ms the rate goes like $[5/1]^3 \sim 125$. The detection of a 10ms VSB is likely to be less efficient than a 1 sec burst. However below $T_{90} - 100$ms increase, giving a factor of 15 increase and giving a short $T_{90}$ enhancement VSB and a possible rate of 2-8/year for BATSE. We have not included the BATSE detection efficiency for VSBs here [8].

**BATSE: Very large + 10 years in space [3$\sigma$]**

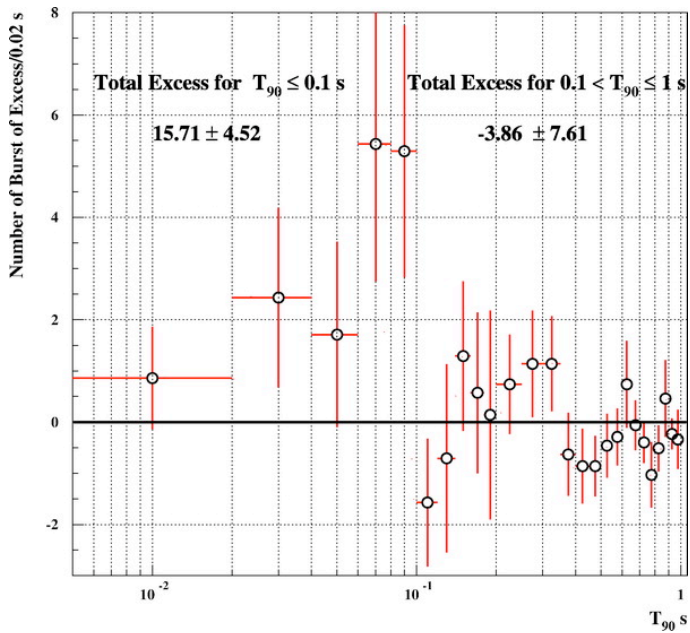

Fig. 1. BATSE GRB events (1991 April 21 2000 May 26). Excess in the GRBs inside the chosen region in the Galactic plane (see Fig. 1) as a function of $T_{90}$. Note that it is likely that the BATSE detection efficiency decreases with decreasing $T_{90}$ (see Ref. 3).

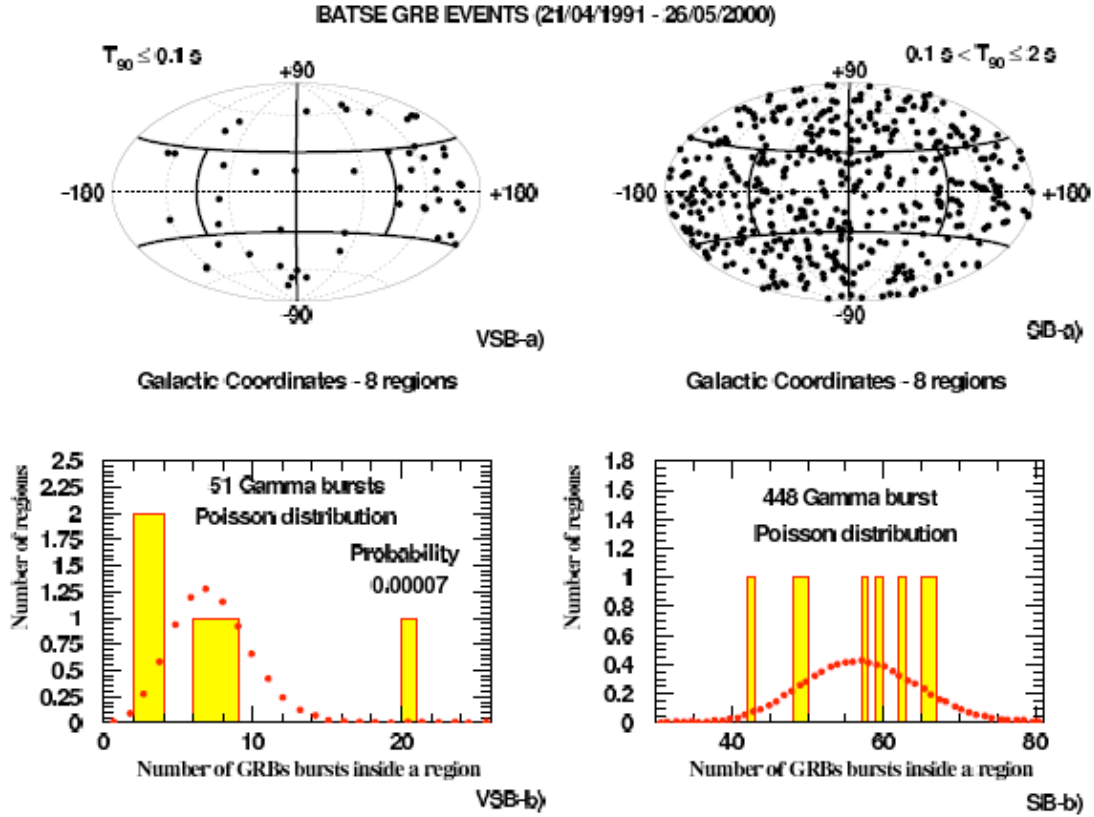

Figure 2. BATSE GRB events (1991 April 4 – 2000 May 26). Angular distribution of the GRBs in Galactic coordinates and the corresponding histograms in comparison with Poisson distribution predictions for two different T90 ranges (filled circles) [2].

## 3. Comparison of the Evidence of Enhancements on the VSB (T90 < 100 ms) for the BATSE, KONUS and Swift Spacecraft Date

We have recently published several papers showing a robust enhancement rate of GRB with T90 < 100 (we define these as VSBs, very short bursts).

(a) BATSE data: Clear asymmetry in the location of VSB showing a likely local (galactic) source, clear evidence for a class of VSB events; see Figure 2 and Figure 1, other effects as well [1].

(b) KONUS data: When studying the energy spectrum of the SGRB we note the VSB class has a much harder group of VSB, see Figure 3 [1][2].

(c) We now turn to a study of the new Swift spacecraft (Fig. 4a) data. There are far fewer events (till July 2009 38 SGRB), 8 events with T90 ≤ 0.1s. We try to fit the simplest linear function; the effect is presented in Figures 4b and the excess VSB result in a 2.7σ difference from the fit in Fig. 4b.

**KONUS: Many years in space, very hard spectrum at short T90 [>3σ]**

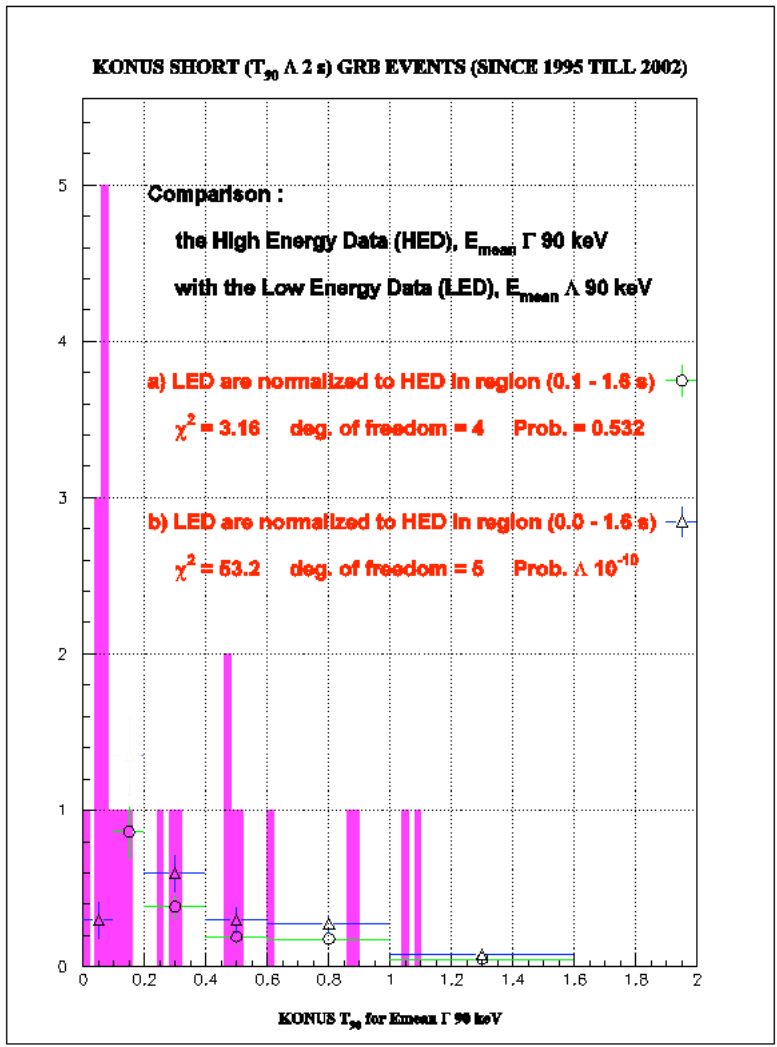

Figure 3. KONUS data with different cuts on the average photon energy $<E_\gamma>$ [1][6].

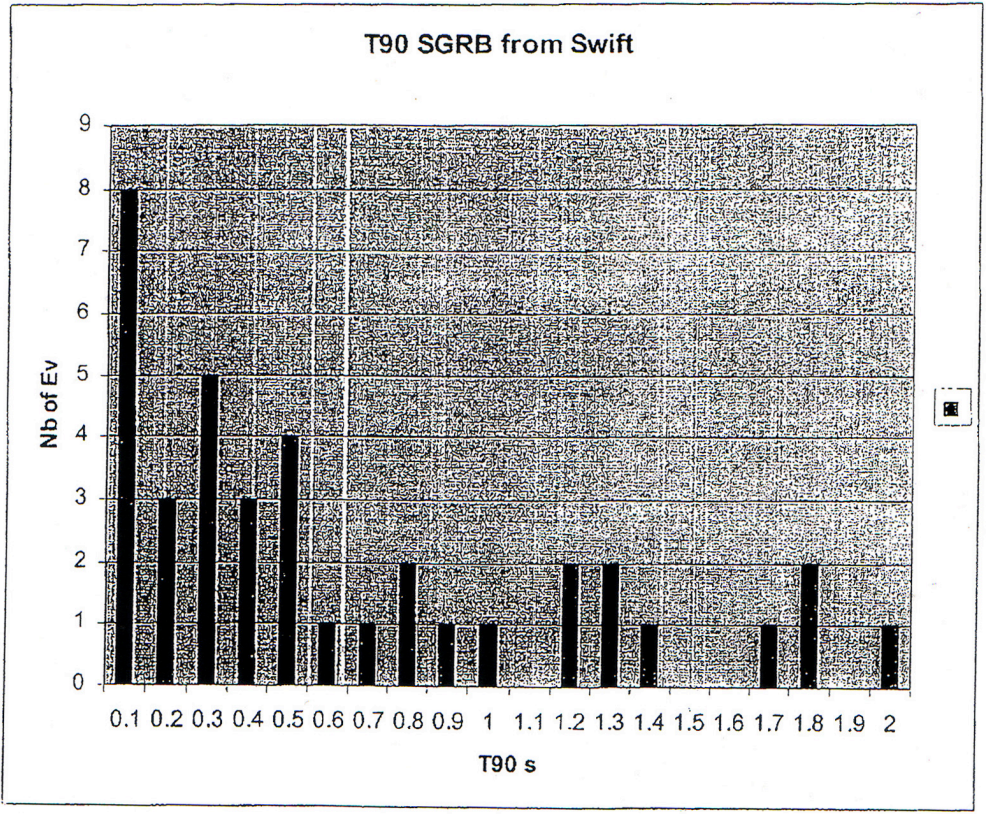

Figure 4a. Swift SGRB data. Swift GRB Table, http://swift.gsfc.nasa.gov/docs/swift/archive/grb_table.html

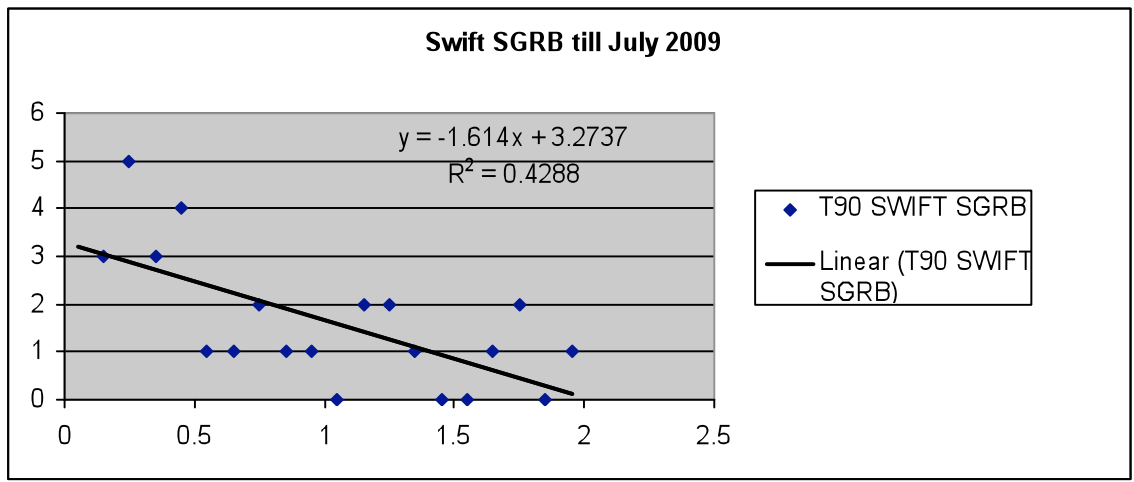

Figure 4b. Linear fit to SWIFT data. The resulting significance of the VSB peak in the swift data $Y(0.05) = 3.193 \Rightarrow \sigma = (8-3.193)/\sqrt{3.193} = 4.807/1.787 = 2.69\sigma$.

We summarize the significance of the VSB excess in Table 3.

Table 3. Significance of the T90 < 100ms Effect: Hypothesis

→ For Swift 2.7σ

→ For BATSE > 3σ

→ For KONUS > 3σ

To take a conservative limit, assume $\sigma_{total} = \sqrt{\sigma_{Swift}^2 + \sigma_{Konus}^2 + \sigma_{BATSE}^2} = \sqrt{22} \cong 4.5\sigma$

Note that 3 experiments have ≥ 2σ. So this is a convincing 4.5σ effect, we believe.

The size of the signal is consistent with local PBH evaporation if we take into account the size of BATSE and the long time of Konus in space.

**4. Time Structure for PBH evaporation**

In order to look for a key signature for PBH existence and evaporation the behavior of the luminosity as a function of time causes a unique time structure (see Table 1). We expect more events at shorter time when the temperature has risen. This effect is best seen where the luminosity of the largest. For evaporation times of less than 100ms there are now many events to study from the BATSE, Konus, Swift spacecraft. To observe the time structure a time duration of 30-50ms is preferred to give a better "lever arm" for the short time enhancement.

In Figure 5 we show the shape of the time structure from the PBH evaporate in the final second. At the time of PBH evaporation of 100ms the black hole temperature is about 10TeV, at the time of 10ms the temperature is 20TeV, and at 1ms we expect 30TeV. This leads to an increasing and very hard photon spectrum at short time that we have previously identified and to an increase of the number of events with decreasing time [1][2][3]. We show here that a number of VSGRBs that exhibit this effect expected for primordial black hole evaporation. In Figure 6a, 6b, and 6c, note that all events have a time signature similar to Figure 5. We use these events as illustrations. There are many more events that have a similar time sequence. For shorter time spreads (14ms) we show a BATSE event that closely shows a time shape signature [4] [Table 4].

At very short times of 10s of ms we expect the detectors detection efficiency to decrease rapidly, cutting off the distribution at the short time side.

Figures 6a, 6b, and 6c show examples of VSB from BATSE that exhibit the time profile of the PBH evaporation (Figure 5). There are many such events but we have studied these in detail using the BATSE TTE data (2μs resolution). We have also fit the T90 value and find in Table 4 [4]:

Table 4. $T_{90}$ from BATSE and from the UCLA fit to BATSE TTE Data [4]

| Event | $T_{90}$ (ms) (BATSE) | $T_{90}$ (UCLA fit) ms |
|---|---|---|
| 2463 | 64 | 49±4.5 |
| 0512 | 183 | 14±0.6 |
| 0432 | 34 | 50±1.8 |

We have also made a Fourier analysis of these events [3][4]. In Figure 7a we show this event blown up with ins time bins. We also show a very approximate curve of the expected time profile for PBH evaporation from Figure 5. They are in excellent agreement.

In Figure 7b we show one of many VSB observed by the KONUS spacecraft. We note that the fine structure again strongly resembles that expected for a PBH (Figure 5). We also show the hardness as a function of time that increases with decreasing time as expected for a PBH evaporation.

We have shown previously that the event hardness increases with decreasing time spread of the VSB, again consistent with PBH evaporation.

Time structure of the Final PBH Evaporation

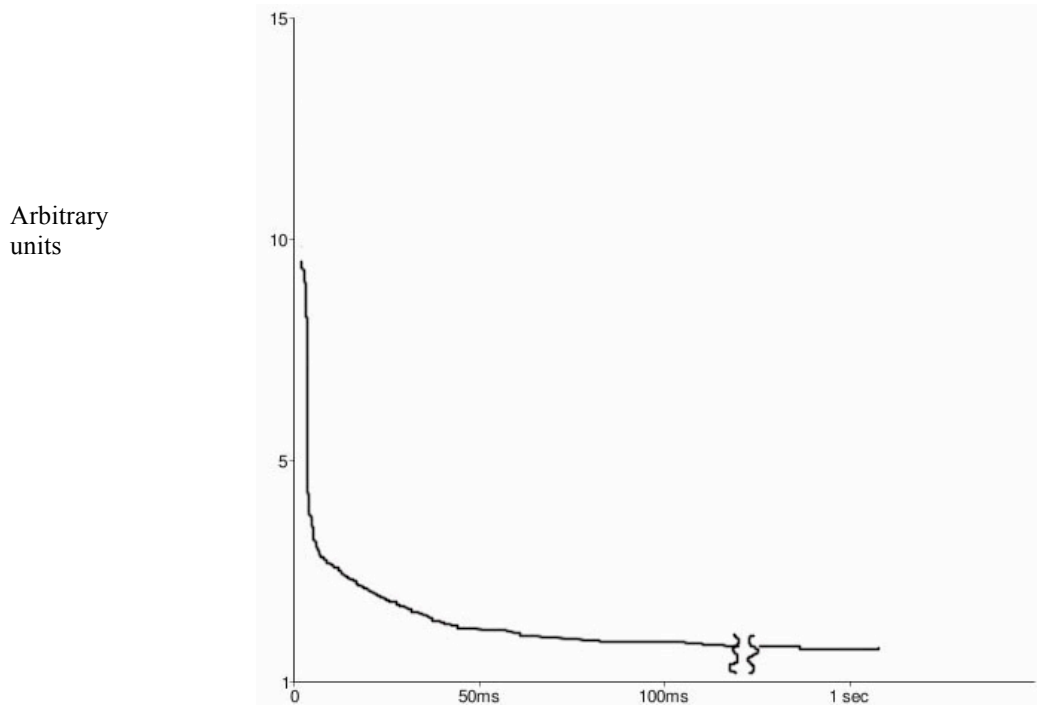

Figure 5. Intensity of evaporation of a PBH in the final stage (from Table 1).

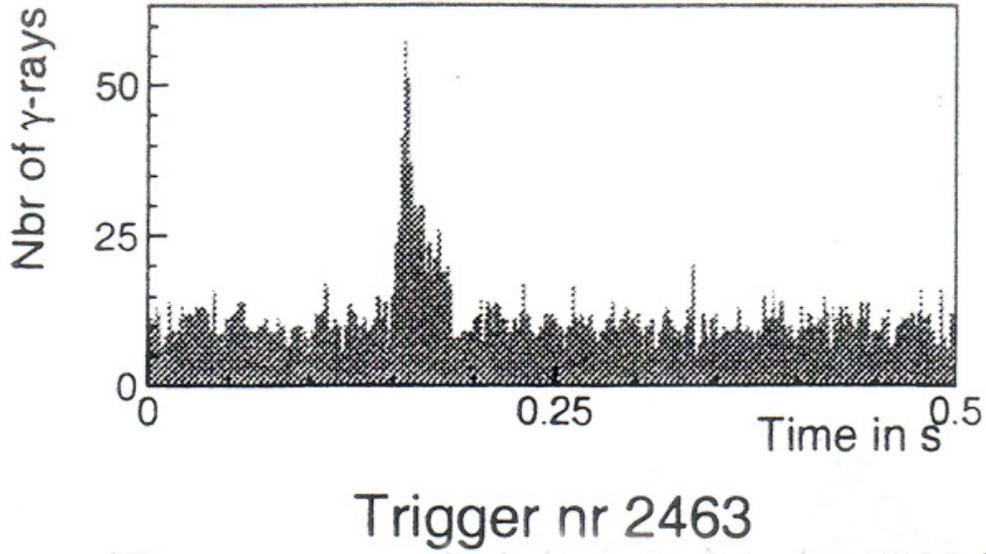

Figure 6a. Time profile of BATSE event 2463, analyzed using the TTE data shown.

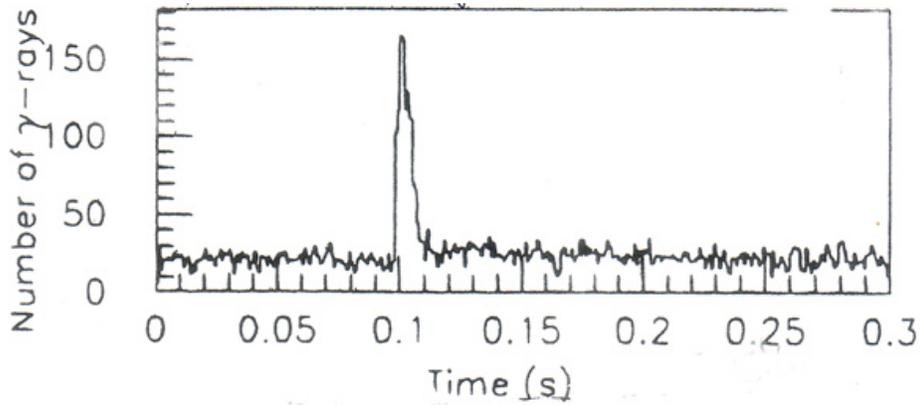

Figure 6b. Time profile of BATSE event 0512 using the TTE data stream.

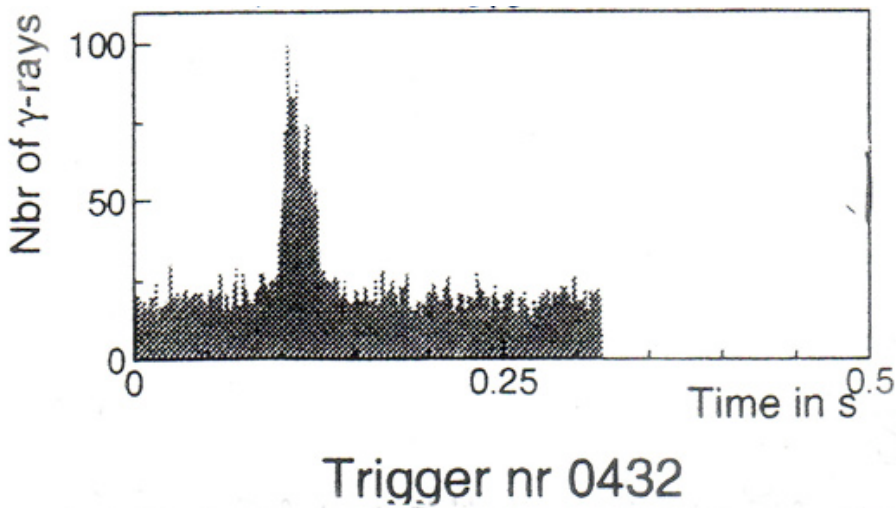

Figure 6c. Time profile of BATSE event 0432 using the TTE data stream.

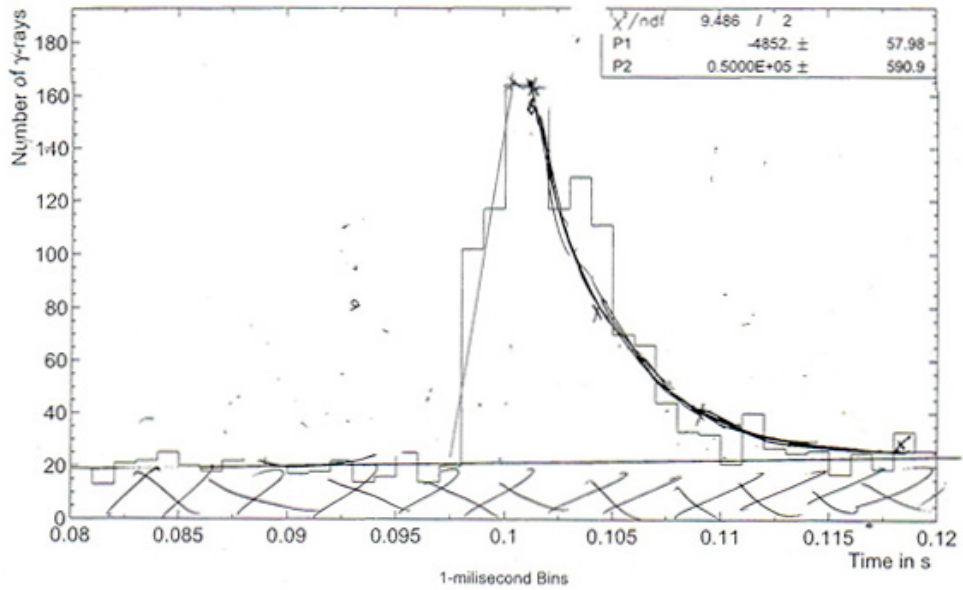

Figure7a. Event 0512 (see Fig. 6b) in an expanded time scale. The approximate shape of the PBH evaporation time profile is shown as an example. The data is from the fit to BATSE TTE data [4].

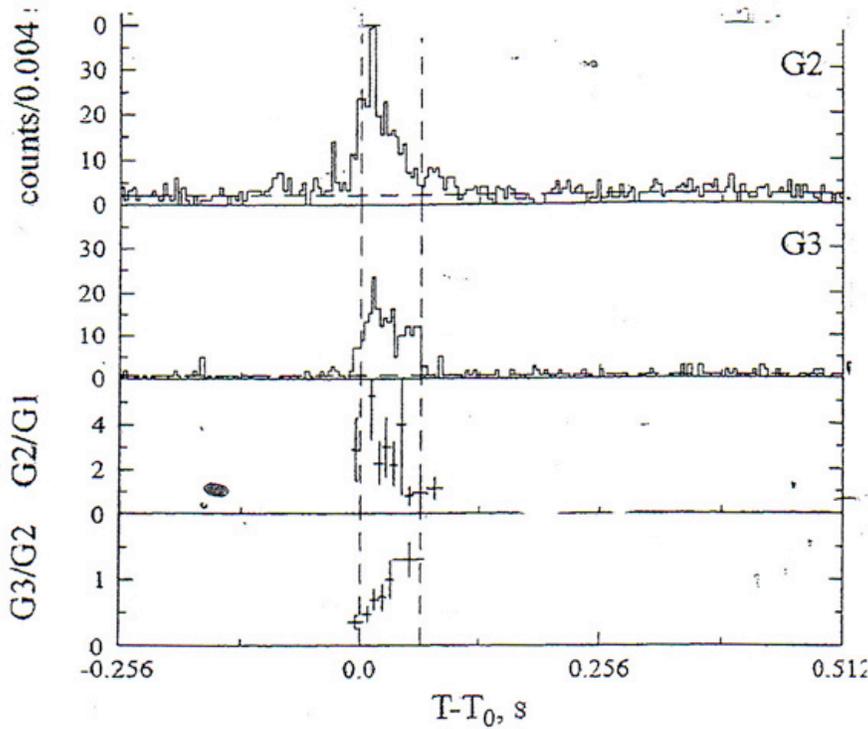

Figure 7b. An event from the Konus spacecraft with $T_{90}$ = 50ms. The shape of the time pulse [G2] and the hardness G2/G1 both show a clear profile similar to that expected from a PBH evaporation (Fig. 5) Table 1.

## 5. Summary

We have shown that the VSB class of gamma ray bursts is very robust in the Swift, BATSE and Konus data. We also show that a number of events exhibit the time structure signature expected for PBH evaporation. We provide references to our earlier work in references 5, 6, 7, 8. Of course the original concept is from Hawking (references 9 and 10). More recent work incorporating quarks and gluons is given in references 11 and 12.

**References**

1. D.B. Cline et al, "Comparison of VSB from BATSE, KONUS and SWIFT", Nuovo Cim. 121B, 1443, 2006; astro-ph/0608158.2.
2. D.B. Cline, C. Matthey, and S. Otwinowski, ApJ 527 (1999), 827.
3. D.B. Cline, C. Matthey and S, Otwinowski, "Evidence for a Galactic origin of Very Short Gamma Ray Bursts and Primordial Black Hole Sources", Astropart. Phys 18(2003), 531; astro-ph/0110276.
4. D.B. Cline, D.A. Sanders, and W.P. Hong, "Further Evidence for some Gamma-Ray Bursts Consistent with Primordial Black Hole Evaporation", ApJ 486 (1997), 169.
5. D.B. Cline, "Some Gamma-Ray Bursts Consistent with Primordial Black Hole Evaporation", ApJ 486 (1997), 169.
6. D.B. Cline et al, "Study of Very Short Gamma Ray Bursts: New Results from BATSE and Konus," ApJ 633 (2005), L.
7. M. Magliocchetti, G. Ghirlanda, A Celotti, "Evidence for Anisotropy in the Distribution of Short-lived Gamma Ray Bursts", MNRAS 343 (2003), 255.
8. D.B. Cline, Nuclear Phys. A 610 (1996), 500c.
9. S.W. Hawking, "Black Hole Explosions?", Nature 248 (1974), 30.
10. B. Carr and S.W. Hawking, "Black Holes in the Early Universe", MNRAS 168 (1974), 399.
11. J.H. MacGibbon and B.J. Carr, "Cosmic Rays from Primordial Black Holes", ApJ 371 (1991), 447.
12. F. Halzen et al, "Gamma Rays and Energetic Particles from Primordial Black Holes", Nature 353 (1991), 897.
13. D.N. Page and S.W. Hawking, " Gamma Ray from Primordial Black Holes", ApJ 206 (1976), 1.